\documentstyle[12pt,preprint,aps,tighten]{revtex}

\begin{document}

\preprint{
\vbox{\hbox{JHU--TIPAC--98006}
      \hbox{hep-ph/9804248}
      \hbox{April 1998} }}

\title{Hybrid Charmonium Production in $B$ Decays}
\author{George Chiladze, Adam F. Falk and Alexey A. Petrov}
\address{
Department of Physics and Astronomy, The Johns Hopkins University\\
3400 North Charles Street, Baltimore, Maryland 21218}
\maketitle
\thispagestyle{empty}
\setcounter{page}{0}
\begin{abstract}%
We study the production of charmonium hybrids in $B$ decays.  We use the operator product expansion and nonrelativistic QCD to organize the various contributions to this process.  We express the decay rate in terms of a few matrix elements which eventually will be fixed by experimental measurements or calculated on the lattice.  While the Fock state expansion is problematic for hybrids, in that there is no perturbative large mass limit, the operator product expansion still provides a model independent framework for phenomenological calculations of hybrid production and decay.  We then use a simple flux tube model to estimate the branching ratio $B\to\psi_g+X$, where $\psi_g$ is a $J^{PC}=0^{+-}$ hybrid, the large production of which could help resolve the low charm multiplicity and semileptonic branching ratio observed in $B$ decays.  We also investigate the possible contributions of tensor charmonium production.  We observe that it is unlikely for either effect to be large enough to play a significant role in resolving these problems.
\end{abstract}

\section{Introduction}

The hybrid states remain an enigmatic feature of hadronic physics. Their existence is predicted in QCD (at least in the large $N$ limit~\cite{cohen}), yet most of their properties have been studied only within models or on the lattice.  This is true even of quarkonium hybrids, because, unlike ordinary quarkonium, hybrids are characterized by excitations of the nonperturbative gluon field.  Even for large mass $m_Q$, there is no reason for a $\overline QQ$ hybrid to be a compact object, and neither the lattice nor models support such a picture.  The absence of a limit in which hybrids are well understood makes a rigorous analysis of properties such as their masses, decays, and production rates difficult if not impossible to achieve.

Yet it is still important to do what one can.  There is experimental evidence that hybrids exist in the light hadron spectrum~\cite{E852}.  As for charmonium hybrids, it has been suggested that they might play an interesting role in nonleptonic $B$ decays, in particular, in resolving the puzzle of the low semileptonic branching fraction . Essentially, the problem is that there is a discrepancy between the CLEO experimental measurement~\cite{CLEO}
\begin{equation}
  {\cal B}_{\rm SL}^{\rm exp} \equiv \frac{\Gamma (B \to X l \nu)}
  {\Gamma_{\rm tot}} = 0.1019 \pm 0.0037
\end{equation}
and theoretical estimates which prefer ${\cal B}_{\rm SL}^{\rm th} > 0.115$~\cite{BSLproblem,BSLreview}. Here
$\Gamma_{\rm tot}=\Gamma(b \to c \bar c + X) + \Gamma(b \to c + X) +
\Gamma(b \to {\rm no\ charm})$ is the total decay rate of the $B$ meson.
Assuming that $\Gamma (B \to X l \nu)$
can be predicted accurately, the simplest resolution is to boost the nonleptonic decay rate, typically either by enhancing $\Gamma(b \to c \bar c + X)$ or $\Gamma(b \to c + X)$ within the Standard Model~\cite{BSLproblem,BSLreview}, or by enhancing $\Gamma(b \to {\rm no\ charm})$ with new physics~\cite{kagan}.  The first option is constrained, however, by the measured charm multiplicity from CLEO~\cite{CLEO},
\begin{equation}
  n_c \equiv 1 + \frac{\Gamma(b \to c \bar c + X)}{\Gamma_{\rm tot}} -
  \frac{\Gamma(b \to {\rm no\ charm})}{\Gamma_{\rm tot}} = 1.12 \pm 0.05\,,
\end{equation}
which apparently cannot accommodate a large rate $\Gamma(b \to c \bar c + X)$.  While new physics remains a possibility, it is certainly more conservative to look for new mechanisms to resolve the problem within the Standard Model.\footnote{Measurements from LEP~\cite{LEP} of ${\cal B}_{\rm SL}^{\rm exp}= 0.1112 \pm 0.0020$ and $n_c=1.20\pm0.07$ are consistent with the Standard Model, so there are experimental issues to be resolved as well.}

One such possibility is the suggestion that a significant fraction of the $b \to c \bar c + X$ transition is seen as charmless $b \to s$ processes~\cite{close}.  If this were true, then the conventional estimate
of ${\cal B} (B \to {\rm no\ open\ charm}) = 0.052 \pm 0.011$ could be boosted to ${\cal B} (B \to {\rm no\ open\ charm}) \approx 0.16$.  The mechanism proposed to account for this involves the production of a charmonium hybrid which, for dynamical reasons, does not decay to open charm.  For example, in some models of hybrids there are ``selection rules'' which suppress decays into mesons with ``similar'' wavefunctions, or the hybrid can have exotic quantum numbers which prohibit its decay into certain hadronic final states.  In this paper, we will investigate the production of hybrid charmonium in $B$ meson decay.  We will not have anything new to say about whether the hybrids under consideration do, in fact, have the desired masses and branching ratios; this is an extremely model-dependent question, and one which ultimately must be answered by experiment.  Rather, we will address the issue of whether, if such hybrids exist, their production rate could be large enough to make the proposed mechanism feasible.  In doing so, we will identify the model independent aspects of the analysis and develop them as fully as possible.

We will begin in Section II with a brief discussion of the properties of hybrids.  Section III contains a general analysis of hybrid charmonium production in $B$ decays, based on the operator product expansion and nonrelativistic QCD.  In Section IV, we will use a simple flux tube model to estimate the matrix elements which govern the production of this state.  For concreteness we will focus on charmonium with $J^{PC}=0^{+-}$, an exotic meson which cannot decay to $D^{(*)}D^{(*)}$.  Sections V contains observations concerning tensor charmonium production, and we summarize in Section VI.

\section{Quarkonium hybrids}

A hybrid meson is a resonance which contains, in addition to a valence quark and antiquark, excited gluonic degrees of freedom.  Some hybrids have exotic quantum  numbers, such as $J^{PC}=0^{+-}$ or $1^{-+}$, which cannot be realized by a pure $\overline QQ$ configuration, and such states, if observed, are clearly hybrids (or four quark states) of some sort.  Other hybrids have conventional quantum numbers, and hence can mix with ordinary mesons; here the distinction between the two is less clear.  The gluonic excitation itself is realized variously in different models. For example, one may introduce into the constituent quark model a constituent gluon which couples to the $\overline QQ$ state to form a hybrid~\cite{const}.  Alternatively, bag models treat the motion of the diquark state in the background of a gluonic standing wave in the cavity~\cite{bag}. Flux tube models realize gluonic degrees of freedom by introducing collective phonon excitations on the color flux tube which connects the quark and antiquark at distances of the order of $\Lambda_{\rm QCD}$~\cite{flux,barnes,closepage}.  Hybrids also have been studied using QCD sum rules~\cite{sumrules} and on the lattice~\cite{lattice,manke}.

Despite the wide variation in these physical pictures, the models of hybrids display some common features.  They typically predict the same quantum numbers for many of the lowest lying states, even if the masses are somewhat different. It also has been noted that the states in these models often possess symmetries which forbid or suppress hybrid decay into two mesons with identical wavefunctions.  This particular property, if true, might prove useful for resolving the semileptonic branching ratio and charm multiplicity problems in $B$ decays, since the production of charmonium hybrids which do not decay to $D^{(*)}D^{(*)}$ could provide channels through which a $c\bar cs$ final state could hadronize and then appear as a final state with no open charm.  Such a mechanism would allow for an enhancement in $\Gamma(b \to c \bar c + X)$, but with no enhancement, in fact a suppression, of $n_c$.

What would be required for this mechanism is one or more narrow hybrid charmonium states favoring light hadronic decay channels.  At this point, the evidence for the existence of such a state is very model dependent, and has in no way yet been confirmed by genuine QCD calculations.  For the purposes of this paper, we will posit that such a state exists and investigate its production in $B$ decays.  To be specific, we shall study the {\it exotic} hybrid state with $J^{PC}=0^{+-}$, where strong decay to $D^{(*)}D^{(*)}$ final states is forbidden by charge conjugation or parity. This state cannot be realized by a $c \bar c$ combination and therefore has a unique signature of a hybrid.  However, the formalism developed here can be applied generally to the study of any hybrid charmonium production in $B$ decays, as well as to hybrid photoproduction and hybrid quarkonium annihilation.

A times, it will be useful for us to have a simple but physical model of hybrid mesons, and for this purpose we will turn to the original flux tube model of Isgur and Paton~\cite{flux}. (This model has been elaborated considerably; see, for example, Ref.~\cite{barnes}.)  In this model, the quark and antiquark are separated by a distance of order $1/\Lambda_{\rm QCD}$ and connected by a string of color flux.  Transverse oscillations are then excited on the string, corresponding to collective degrees of freedom of the glue.  The excitation state of the string governs the effective potential of the $Q$ and $\overline Q$.  For example, with one phonon in the lowest excited state, 
the potential energy as a function of the separation $r$ is
\begin{equation}\label{ipv}
   V(r)=-{4\alpha_s\over3r}+c+br+{\pi\over r}
   \left[1-\exp\left(-fb^{1/2}r\right)\right]\,,
\end{equation}
where $b\simeq 0.18\,{\rm GeV}^2$ is the string tension, $c\simeq-0.7\,{\rm GeV}^2$ is a constant, and $f$ is a parameter of order one which governs the $r\to0$ behavior of the flux tube.  Already in this sector, a rich spectrum of hybrid charmonium states is predicted, with $J^{PC}=0^{\pm\mp}$, $1^{\pm\mp}$, $2^{\pm\mp}$ and $1^{\pm\pm}$.  Of these, the $0^{+-}$ and $1^{-+}$ states are exotic.  The $CP$ quantum numbers of these hybrids are related to the total spin $S$ of the quark and antiquark by $\eta_C\eta_P=(-1)^S$, as compared to $\eta_C\eta_P=(-1)^{S+1}$ for ordinary mesons.  Thus in this model the hybrids can mix with conventional states only via hyperfine transitions.

Hybrids are different from ordinary quarkonium in an important respect which makes their study problematic, which is that they have no simple limit as $m_Q\to\infty$.  In this limit, ordinary quarkonium admits a controlled Fock state decomposition, in which the leading term is a $\overline QQ$ in a color singlet with fixed quantum numbers $^{2S+1}L_J$.  Higher order terms, such as color octet configurations, are suppressed by powers of $\alpha_s(m_Q)\sim v$, where $v$ is the relative velocity of the $Q$ and the $\overline Q$.  The expansion is possible because for large $m_Q$ the quarkonium is a small state, of size $(m_Qv)^{-1}$, whose interaction with the color field is governed by a multipole expansion.  By contrast, the minimum of the potential (\ref{ipv}) is at a separation $r_0$ of order $1/\Lambda_{\rm QCD}$, independent of $m_Q$.  For large $m_Q$, fluctuations about $r_0$ are small, but the state itself is not compact.\footnote{One also might imagine quarkonium hybrids with the structure of a small color singlet quarkonium bound to a large glueball.  There is little evidence at present for such bound states, and we will not consider them here.}  This appears to be a generic feature of hybrid models; for example, in a constituent gluon model, the $Q$ and $\overline Q$ are in a color octet and repel each other at short distances.  At large $m_Q$, one expects a situation somewhat like a heavy rigid rotor, with nearly degenerate rotational bands of states.  This behavior has been observed in lattice studies~\cite{manke}. 

The absence of a controlled Fock state expansion for hybrid quarkonium can be understood from simple physical considerations. In the case of ordinary heavy quarkonium, a soft glue configuration
has a small overlap with the compact two-quark
state, since the Compton wavelength of the soft gluons, of the order of $\Lambda_{QCD}^{-1}$, is much larger than the distance between the quarks. By contrast, the size of the heavy hybrid state remains finite, of
order $\Lambda_{QCD}^{-1}$, in the heavy quark limit,
thus allowing for a significant interaction with the
soft gluonic modes. Yet, in this limit heavy quarks still become nonrelativistic, and thus the production of virtual heavy quark pairs is strongly suppressed.

Nevertheless, there is still a sort of Fock state expansion for quarkonium hybrids.  Although the gluonic degrees of freedom are intrinsically nonperturbative, we can organize the various components according to the quantum numbers of the $Q$ and the $\overline Q$.  For example, for the $0^{+-}$ charmonium hybrid state which we will study later, we can write a decomposition of the form
\begin{equation} \label{fock}
  | \psi_g \rangle = A | c \bar c(^3S_1)_{1,8} + g_1 \rangle +
  B | c \bar c(^3P_J)_{1,8} + g_2 \rangle +
  C | c \bar c(^1S_0)_{1,8} + g_3 \rangle + \dots\,,
\end{equation}
where the subscripts $1,8$ indicate the color of the $\bar cc$ pair and the $g_i$ are various gluonic configurations.  While one is always free to do this, here there is no model-independent hierarchy among the coefficients.  This is related to the lack of a unique $\bar cc$ ``baseline'' state, and the lack of a multipole expansion to govern transitions between the different Fock components.  (Note that heavy quark spin symmetry still governs hyperfine transitions, such as between $\bar cc(^3S_1)_{1,8}$ and $\bar cc(^1S_0)_{1,8}$.)  Nonetheless, {\it within particular models\/} it is typically the case that a given hybrid is dominated by one or two $\bar cc$ configurations.  For large $m_c$, the $c$ and the $\bar c$ move nonrelativistically  in the potential $V(r)$, and an expansion in powers of the velocity $v$ is available.  Hence it is useful to organize our calculation of hybrid production in $B$ decays with a decomposition such as (\ref{fock}) in mind.

\section{Hybrid quarkonium production in $B$ decay}

\subsection{Factorization}

We turn now to the calculation of hybrid charmonium production in
$B$ decays, using the formalism of nonrelativistic QCD (NRQCD), an effective field theory useful for describing nonrelativistic degrees of freedom~\cite{nrqcd}.  The applicability of this technique rests on the fact that even though a hybrid is not necessarily small as $m_Q\to\infty$, the heavy quarks are still moving slowly and virtual $\overline QQ$ production is suppressed.  In the usual way, we will first calculate the production rate in the framework of full QCD, separating hard and soft degrees of freedom.  Then we will compute the rate in NRQCD, and use the perturbative calculation to match the coefficients of the NRQCD matrix elements.

We begin by writing the inclusive $B$ meson decay rate to a (hybrid) charmonium state, $B\to\psi_g+X$, in the $B$ rest frame
\begin{equation} \label{rate1}
  \Gamma^{\rm QCD} = \frac{1}{2 m_b}
  \int \frac{d^4 P}{(2 \pi)^4}
  \prod_{i=1}^{N-1} \int \frac{d^3 k_i}{(2 \pi)^3 2 E_i}
  (2 \pi)^4 \delta^{(4)} \Bigl(p_B - P - \sum_i k_i\Bigr)\, 
  |\langle \psi_g + X |{\cal H}_{eff}|B \rangle|^2\,,
\end{equation}
where $P$ is the momentum of the $\psi_g$ and $k_i$ are the momenta of the other particles in the final state.  Note that we keep the $\psi_g$ off mass shell for now; we will impose this condition later, when we match onto NRQCD.  The effective Hamiltonian takes the form
\begin{eqnarray} \label{Heff1}
  {\cal H}_{eff} &=& \frac{G_F}{\sqrt{2}} V_{cb} V_{cs}^*
  \left [ C_1(\mu) O_1(\mu) + C_2(\mu) O_2(\mu) \right ], \nonumber \\
  O_1 &=& \bar s_i \gamma_\mu (1 + \gamma_5) c^j ~
  \bar c_j \gamma^\mu (1 + \gamma_5) b^i, \nonumber \\
  O_2 &=& \bar s_i \gamma_\mu (1 + \gamma_5) c^i ~
  \bar c_j \gamma^\mu (1 + \gamma_5) b^j,
\end{eqnarray}
where $C_1(m_b) \simeq -0.29$, $C_2(m_b) \simeq 1.1$, and we have neglected small contributions from penguin operators.  Note that the scale dependence of the operators is compensated by the scale dependence of the Wilson coefficients, which we will no longer make explicit. Using Fierz identities for the color matrices, Eq.~(\ref{Heff1}) may be rewritten to bring it to a form more suitable for our calculation,
\begin{eqnarray} \label{Heff2}
  {\cal H}_{eff} &=& \frac{G_F}{\sqrt{2}} V_{cb} V_{cs}^*
  \Biggl [ \left ( C_1 + \textstyle{1\over3}C_2 \right)
  \bar c \gamma_\mu (1 + \gamma_5) c ~\bar s \gamma^\mu (1 + \gamma_5) b
  \nonumber \\
  &&\qquad\qquad\quad \mbox{}+ 2 C_2\, \bar c \gamma_\mu 
  (1 + \gamma_5) T^a c ~\bar s \gamma^\mu 
  (1 + \gamma_5) T^a b\Biggr ],
\end{eqnarray}
with $T^a=\lambda^a/2$. Here we are interested in the production of the hybrid charmonium states, in which there is no preference for the color singlet $\bar cc$ combination.  Hence the first term in Eq.~(\ref{Heff2}) may be considered as a perturbation of the second, since $( C_1 + {1\over3}C_2)\ll C_2$, and we will neglect it.  Defining $\Gamma_\mu = \gamma_\mu (1 + \gamma_5)$, we may write
\begin{eqnarray}
  &&\prod_{i=1}^{N-1} \int \frac{d^3 k_i}{(2 \pi)^3 2 E_i}
  (2 \pi)^4 \delta^{(4)} \Bigl(p_B - P - \sum_i k_i\Bigr)\, 
  |\langle \psi_g + X |{\cal H}_{eff}|B \rangle|^2 \nonumber\\
  &&\qquad\qquad\qquad\qquad=
  2 G_F^2 \left |  
  V_{cb} V_{cs}^* \right |^2 C_2^2 
  \,\,\,\int\!\!\!\!\!\!\!\!\!\!\sum_{X_1,X_2}
  \langle B |\bar b \Gamma_\mu T^a s |X_1 \rangle
  \langle X_1 |\bar s \Gamma_\nu T^b b|B \rangle \nonumber \\
  &&\qquad\qquad\qquad\qquad\qquad\qquad\qquad\qquad\qquad\times 
  \langle 0 |\bar c \Gamma^\mu T^a c |\psi_g + X_2 \rangle
  \langle \psi_g + X_2 |\bar c \Gamma^\nu T^b c|0 \rangle \nonumber\\
  &&\qquad\qquad\qquad\qquad= 
  2 G_F^2 \left | V_{cb} V_{cs}^* \right |^2 C_2^2 ~B^{ab}_{\mu \nu}
  C_{ab}^{\mu \nu}\,. 
\end{eqnarray}
Using the optical theorem, and the fact that all asymptotic states are color neutral, $B^{ab}_{\mu \nu}$ may be related to the imaginary part of a time-ordered product of color singlet currents,
\begin{equation} \label{btens}
  B^{ab}_{\mu \nu} = \frac{\delta^{ab}}{6}\, {\rm Im} \,
  i \int d^4 y\, e^{i P y}\, \langle B | T\,\{j^\dagger_\mu(y), j_\nu(0)
  \} |B \rangle\,.
\end{equation}
This expression for $B^{ab}_{\mu\nu}$ could be a starting point for the usual calculation of $1/m_b$ corrections.  The time ordered product (\ref{btens}) may be expanded in inverse powers of $m_b$, yielding a tower of operators whose expectation values in the $B$ meson are organized according to the heavy quark expansion~\cite{inclusive}.  Here we will concern ourselves only with $1/m_c$ corrections and use the lowest order result, which coincides with the parton model expression for $B^{ab}_{\mu \nu}$,
\begin{eqnarray}
  B^{ab}_{\mu \nu} &=& {4 \pi m_b^2\over3}\, \delta^{ab}\, \delta((p_b-P)^2)
  \Bigg[ - g_{\mu \nu} \left(1- \frac{1}{m_b^2} p_b \cdot P \right )
  +\frac{2}{m_b^2} p_{b \mu} p_{b \nu} \nonumber \\
  &&\qquad\qquad\qquad\qquad\qquad\mbox{}-
   \frac{1}{m_b^2} \left (P_\mu p_{b \nu} + P_\nu p_{b \mu} \right)
  + \frac{i}{m_b^2} \epsilon_{\mu \nu \alpha \beta} P^\alpha p_b^\beta
  \Bigg ]\,,
\end{eqnarray}
where we neglect the mass of the $s$ quark.  The leading nonperturbative corrections to this result would be proportional to $\lambda_{1,2}/m_b^2$.  Neglecting these small effects, the decay rate in full QCD is then written as
\begin{equation} \label{gqcd}
  \Gamma^{\rm QCD} = \frac{N}{2 m_b}
  \int \frac{d^4 P}{(2 \pi)^4}
  B^{ab}_{\mu \nu} C_{ab}^{\mu \nu}
\end{equation}
with $N=2 G_F^2 \left | V_{cb} V_{cs}^* \right |^2 C_2^2$.  The expression
of Eq.~(\ref{gqcd}) must then be matched with the equivalent expression in NRQCD,
\begin{equation} \label{gnrqcd}
  \Gamma^{\rm NRQCD} = \frac{N}{2 m_b}
  \int \frac{d^4 P}{(2 \pi)^4}
  \sum_n \frac{C_n}{m_c^{d_n-4}}\, \langle O_n^{\psi_g} \rangle\,
  2\pi\delta(P^2-m_\psi^2),
\end{equation}
where $O_n^{\psi_g}$ represents a set of NRQCD operators of increasing dimension $d_n$, and we have extracted the on-shell $\delta$ function explicitly.  We are unable to calculate matrix elements of these operators in a model independent fashion, so as usual they are left as free parameters. Note that the same matrix elements would govern the total annihilation width of the hybrid or the hybrid photoproduction cross section.  Eventually, one might hope to extract the leading matrix elements from a small number of experiments.

\subsection{Perturbative QCD Calculation}

We now turn to the perturbative calculation of $C_{ab}^{\mu \nu}$,
\begin{equation}
  C_{ab}^{\mu \nu} = \langle 0 |\bar c \Gamma^\mu T^a c |\psi_g + X_2
   \rangle
  \langle \psi_g + X_2 |\bar c \Gamma^\nu T^b c|0 \rangle =
  \langle 0 |\bar c \Gamma^\mu T^a c\, P_{\bar cc}\, \bar c \Gamma^\nu T^b c
  |0\rangle,
\end{equation}
where we have introduced the projection operator $P_{\bar cc}=|\bar cc+X_3\rangle\,\langle\bar cc+X_3|$ onto the desired $\bar cc$ state within the $\psi_g$.  The relativistic four-component spinors $u_c$ may be boosted to the $\bar cc$ center of mass frame and then replaced by nonrelativistic two-component spinors $\xi$ and $\eta$, using the identities\footnote{Our definition of $\gamma_5$ differs by a sign from that of Ref.~\cite{nrqcd}.}~\cite{nrqcd}
\begin{eqnarray}
  \bar u_c \gamma_\mu T^a u_c &=& \Lambda_j^\mu 
  \left [ 2 E_c \xi^\dagger \sigma^j   T^a
  \eta - \frac{2}{E_c + m_c} q^j \xi^\dagger ({\bf q \cdot \sigma}) T^a \eta   
  \right ],
  \nonumber \\
  \bar u_c \gamma_\mu \gamma_5 T^a u_c 
  &=& - \frac{m_c}{E_c}\, P^\mu \xi^\dagger T^a   
  \eta +
  2 i \Lambda_j^\mu \xi^\dagger ({\bf q \times \sigma})^j T^a \eta,
\end{eqnarray}
where $\Lambda_j^\mu$ is the Lorentz boost matrix.  For clarity, we introduce primes for one of the pairs of spinors.  Keeping terms up to order ${\bf q}^2$, we obtain at lowest order in QCD perturbation theory,
\begin{eqnarray} \label{ctens}
  C_{ab}^{\mu \nu} &=& 4 m_c^2 \Lambda_j^\mu \Lambda_k^\nu \Biggl \{
  \left [1 + \left (\frac{{\bf q'}^2}{2 m_c^2} + \frac{{\bf q}^2}
  {2 m_c^2}\right ) \right ]
  {\xi'}^\dagger \sigma^j T^a \eta' \eta^\dagger \sigma^k T^b \xi
  \nonumber \\
  &&\qquad\qquad\quad\mbox{}- \frac{1}{2 m_c^2} \left [
  {\xi'}^\dagger {\bf q'}^j ({\bf q' \cdot \sigma})  T^a \eta'
  \eta^\dagger \sigma^k T^b \xi +
  {\xi'}^\dagger \sigma^j T^a \eta'
  \eta^\dagger {\bf q}^k ({\bf q \cdot \sigma}) T^b \xi \right ]
  \nonumber \\
  &&\qquad\qquad\quad\mbox{}+
  \frac{1}{m_c^2} {\xi'}^\dagger ({\bf q' \times \sigma})^j T^a 
  \eta'
  \eta^\dagger ({\bf q \times \sigma})^k T^b \xi \Biggr \}
  \\
  &&\mbox{}+P^\mu P^\nu \left [ 1 - \left (\frac{{\bf q'}^2}
  {2 m_c^2} +\frac{{\bf q}^2}{2 m_c^2} \right ) \right ]
  {\xi'}^\dagger T^a \eta' \eta^\dagger T^b \xi + \dots\,, 
  \nonumber
\end{eqnarray}
where there is an implicit projection onto the desired spin and angular momentum configuration of the quarks.  Note that Eq.~(\ref{ctens}), along with (\ref{btens}) and (\ref{gqcd}), is valid for {\it any} charmonium hybrid state. If we now specify to the case of unpolarized hybrids, we arrive at a simple expression for $\Gamma^{\rm QCD}$,
\begin{eqnarray} \label{finalqcd}
  \Gamma^{\rm QCD} &=& {N\over 2m_b}
  \int \frac{d^4 P}{(2 \pi)^4}\frac{2 \pi m_b^2}{9\mu_\psi}
  (1- \mu_\psi)(1+2 \mu_\psi)\delta((p_b-P)^2)\times 4m_c^2\nonumber \\
  &&\quad\times \Biggl\{
  \left (1 + \frac{{\bf q'}^2+{\bf q}^2}
  {2 m_c^2} \right )
  {\xi'}^\dagger {\bf \sigma}^j T^a \eta' 
  \eta^\dagger {\bf \sigma}^j T^a \xi+
  \frac{1}{m_c^2} {\xi'}^\dagger ({\bf q' \times \sigma})^j T^a \eta'
  \eta^\dagger ({\bf q \times \sigma})^j T^a \xi   \nonumber\\
  &&\qquad\qquad\mbox{}+
  \frac{3 \mu_\psi}{4 \mu_c (1+2 \mu_\psi)}
  \left(1 - \frac{{\bf q'}^2+{\bf q}^2}
  {2 m_c^2} \right)
  {\xi'}^\dagger T^a \eta'\eta^\dagger T^a \xi\\
  &&\qquad\qquad\mbox{}-{1\over2m_c^2}\left [
  {\xi'}^\dagger {\bf q'}^j ({\bf q' \cdot \sigma})  T^a \eta'
  \eta^\dagger \sigma^j T^a \xi +
  {\xi'}^\dagger \sigma^j T^a \eta'
  \eta^\dagger {\bf q}^j ({\bf q \cdot \sigma}) T^a \xi \right ]
  + \dots\Biggr\}\,,\nonumber
\end{eqnarray}
where $\mu_c=m_c^2/m_b^2$ and $\mu_\psi=P^2/m_b^2$, in anticipation of matching onto the hybrid state for which $P^2=m_\psi^2$.  Eq.~(\ref{finalqcd}) expresses the decay rate in terms of matrix elements of operators of increasing dimension, and the projection onto an unpolarized hybrid is understood.  The explicit factor of $4m_c^2$ comes from the original normalization of the relativistic $\bar cc$ state.

\subsection{NRQCD Calculation}

To perform the matching, we have to calculate $\Gamma (B \to \psi_g + X)$ in perturbative NRQCD. The expansion takes the form
\begin{eqnarray} \label{ope}
\Gamma^{\rm NRQCD} &=& \frac{N}{2 m_b}
  \int \frac{d^3 P}{(2 \pi)^3 2 E_\psi}\,2m_\psi\,
  \Biggl [
  C_8^1(^3S_1)\langle O_8^{\psi_g}(^3S_1) \rangle +
  C_8^1(^1S_0) \langle O_8^{\psi_g}(^1S_0) \rangle
  \nonumber \\
  &&\mbox{}+ 
  \frac{C_8^1(^3P_1)}{m_c^2} \langle O_8^{\psi_g}(^3P_1) \rangle +
  \frac{C_8^2(^3S_1)}{m_c^2} \langle P_8^{\psi_g}(^3S_1) \rangle +
  \frac{C_8^2(^1S_0)}{m_c^2} \langle P_8^{\psi_g}(^1S_0) \rangle
  \nonumber \\
  &&\mbox{}+  
  \frac{C_8^3(^3S_1)}{m_c^2} \langle Q_8^{\psi_g}(^3S_1) \rangle +\dots
  \Biggr ],
\end{eqnarray}
where the NRQCD operators are given by
\begin{eqnarray}\label{operators}
  \langle O_8^{\psi_g}(^3S_1) \rangle &=&
  \langle 0| \chi^\dagger {\bf \sigma}^j T^a \psi P_{\psi_g}
  \psi^\dagger {\bf \sigma}^j T^a \chi |0 \rangle,
  \nonumber \\
  \langle O_8^{\psi_g}(^1S_0) \rangle &=&
  \langle 0| \chi^\dagger T^a \psi P_{\psi_g}
  \psi^\dagger T^a \chi |0 \rangle,
  \nonumber \\
  \langle O_8^{\psi_g}(^3P_1) \rangle &=&
  \frac{1}{2} \langle 0| \chi^\dagger
  ( -\textstyle{i\over2}{\bf D}  {\bf \times \sigma})^j T^a \psi
  P_{\psi_g} \psi^\dagger
  ( -\textstyle{i\over2}{\bf D}  {\bf \times \sigma})^j T^a \chi 
  |0\rangle,
  \nonumber\\
  \langle P_8^{\psi_g}(^3S_1) \rangle &=&
  \frac{1}{2} \left [
  \langle 0| \chi^\dagger {\bf \sigma}^j
  ( -\textstyle{i\over2} {\bf D})^2 T^a \psi P_{\psi_g}
  \psi^\dagger {\bf \sigma}^j T^a \chi |0 \rangle
  +\langle 0|\chi^\dagger {\bf \sigma}^j T^a \psi P_{\psi_g}
  \psi^\dagger {\bf \sigma}^j
  ( -\textstyle{i\over2}{\bf D} )^2 T^a \chi|0 \rangle \right ],
  \nonumber \\
  \langle P_8^{\psi_g}(^1S_0) \rangle &=&
  \frac{1}{2} \left[
  \langle 0| \chi^\dagger
  ( -\textstyle{i\over2}{\bf D}  )^2 T^a \psi P_{\psi_g}
  \psi^\dagger T^a \chi |0 \rangle +
  \langle 0| \chi^\dagger T^a \psi P_{\psi_g}
  \psi^\dagger
  ( -\textstyle{i\over2}{\bf D}  )^2 T^a \chi |0 \rangle
  \right ],
  \nonumber \\
  \langle Q_8^{\psi_g}(^3S_1) \rangle &=&
  \frac{1}{2} \left[
  \langle 0| \chi^\dagger
  ( -\textstyle{i\over2}{\bf D}  )^j
  ( -\textstyle{i\over2}{\bf D}\cdot\sigma  )T^a \psi P_{\psi_g}
  \psi^\dagger \sigma^j T^a \chi |0 \rangle \right.\nonumber\\
  &&\qquad\qquad\left. 
  \mbox{}+\langle 0| \chi^\dagger \sigma^jT^a \psi P_{\psi_g}
  \psi^\dagger
  ( -\textstyle{i\over2}{\bf D}  )^j
  ( -\textstyle{i\over2}{\bf D}\cdot\sigma  ) T^a \chi |0 \rangle
  \right ].
\end{eqnarray}
Here $P_{\psi_g}$ is a projection onto the hybrid state $\psi_g$, and ${\bf D}=\roarrow{\bf D}-\loarrow{\bf D}$.  By convention, in Eq.~(\ref{operators}) the state $|\psi_g\rangle$ is normalized nonrelativistically.  This is made explicit by the prefactor $2m_\psi$, so the coefficients $C_i$ remain dimensionless, as they are in Eq.~(\ref{gnrqcd}).  Color singlet operators could also be included.  Note that each derivative insertion brings a power of the velocity of the heavy quark, $p/m_c \sim v$. For instance, $O_8^{\psi_g}(^3P_1)$ is suppressed compared to $O_8^{\psi_g}(^3S_1)$ by $v^2$.  Nonrelativistic kinematics clearly is crucial for controlling the number of nonperturbative parameters, and it is therefore not practical to extend these calculations to the case of lighter hybrids.

It is now straightforward to evaluate the NRQCD coefficients, by matching full NRQCD states to perturbative ones.  We have kept the normalizations of the states explicit so that we can treat them carefully.  With our conventions, we must match quantities in the manner
\begin{equation}
  4m_c^2\,{\xi'}^\dagger T^a \eta'\eta^\dagger T^a \xi\longleftrightarrow
  2m_\psi\,\langle 0| \chi^\dagger T^a \psi P_{\psi_g}
  \psi^\dagger T^a \chi |0 \rangle\cdot2\pi\delta(P^2-m_\psi^2)\,,
\end{equation}
where both sides have mass dimension two.  We find
\begin{eqnarray}
  C_8^1(^3S_1) &=& {2\pi\,m_b^2\over9\mu_\psi}\,
  (1- \mu_\psi)(1+2 \mu_\psi) \delta((p_b-P)^2),
  \nonumber \\
  C_8^1(^1S_0) &=& \frac{3 \mu_\psi}{4 \mu_c (1+2 \mu_\psi)}\, C_8^1(^3S_1),
  \nonumber \\
  C_8^1(^3P_1) &=& 2C_8^1(^3S_1),
  \nonumber \\
  C_8^2(^3S_1) &=& C_8^1(^3S_1),
  \nonumber \\
  C_8^2(^1S_0) &=& -C_8^1(^1S_0).
  \nonumber \\
  C_8^3(^3S_1) &=& -C_8^1(^3S_1),
\end{eqnarray}
As discussed in Section II, the Fock space decomposition of hybrid states is not controlled by a multipole expansion.  However, the number of matrix elements important for production processes still can be reduced by keeping those that are leading in powers of $1/m_c$, or equivalently, in powers of $v$.  For instance, in this case only two matrix elements appear at leading order.  (We remind the reader that in the case of hybrids, the identification of the leading matrix contributions is unavoidably model dependent.)

We can now perform the remaining integrations to obtain the partial decay rate for $B\to \psi_g+X$,
\begin{eqnarray} \label{answer}
  \Gamma &=&\Gamma_0\,
  {4(1-\mu_\psi)^2(1 + 2 \mu_\psi )\over m_b^2 m_\psi}
   \Biggl [\langle O_8^{\psi_g}(^3S_1) \rangle
  + \frac{3 \mu_\psi}{4 \mu_c (1+2 \mu_\psi)}
  \langle O_8^{\psi_g}(^1S_0) \rangle
  +{2\over m_c^2} \langle O_8^{\psi_g}(^3P_1) \rangle \nonumber\\
  &&\qquad\quad\mbox{}
  +{1\over m_c^2} \langle P_8^{\psi_g}(^3S_1) \rangle
  -\frac{3 \mu_\psi}{4 \mu_c (1+2 \mu_\psi)m_c^2}
  \langle P_8^{\psi_g}(^1S_0) \rangle
  -{1\over m_c^2} \langle Q_8^{\psi_g}(^3S_1) \rangle+\dots
  \Biggr ].
\end{eqnarray}
where we have defined $\Gamma_0 =
C_2^2|V_{cb} V_{cs}^*|^2 G_F^2m_b^5/144\pi$, which is the free $b$ quark decay rate induced by the second term in the Hamiltonian (\ref{Heff2}), for $m_c=0$.  Note that as $m_b\to\infty$ for $m_c$ fixed, the branching fraction for $B\to\psi_g+X$ falls as $1/m_b^2$.  The terms proportional to $\langle O_8(^3S_1) \rangle$ and $\langle O_8(^3P_1) \rangle$ agree with the corresponding terms in Ref.~\cite{BBYL}, in which the production of ordinary $P$ wave charmonium in $B$ decays is analyzed.  The formula (\ref{answer}) determines the semi-inclusive decay rate of $B$ into a given hybrid charmonium state in terms of parameters which must be fixed by other experiments. We emphasize that these parameters are {\it universal\/}; for example, the photoproduction cross section will be expressed in terms of the same matrix elements, which depend on the static, equilibrium properties of the hybrids.

\section{Matrix elements in a flux tube model}

Now that we have a general formalism for the production of hybrid charmonium in $B$ decays, we would like to return to the question of whether such a mechanism can account for the ``missing charm puzzle'' and low semileptonic branching ratio of the $B$.  For this purpose, we will focus specifically on the  $J^{PC}=0^{+-}$ hybrid, an exotic meson which cannot decay to $D^{(*)}D^{(*)}$.  Although dynamics may forbid or suppress the $DD$ decay channel of some conventional hybrids~\cite{closepage}, the effect is model dependent, and we will restrict ourselves to this more robust case in which the decays are prohibited by spacetime symmetries.  Of course, it is quite possible for the $0^{+-}$ hybrid to be heavy enough that it decays to multibody states with open charm, since its mass is not known model independently.  However, we will assume for the sake of argument that it is not so massive.

We will estimate the matrix elements (\ref{operators}) in the flux tube model of Isgur and Paton~\cite{flux}.  We are forced to use phenomenological models  because of the lack of a first principles understanding of the structure of hybrid mesons.  Eventually, these matrix elements should be computed on the lattice.  Our model dependent calculation serves two purposes.  First, it allows us to use the expansion (\ref{answer}) to obtain a rough estimate of the magnitude of hybrid charmonium production in $B$ decays.  Second, we can explore the extent to which, within the flux tube model, the expansion is well behaved.  We will use the flux tube model in its least elaborate form, and with some additional assumptions, on the grounds that this approach is sufficient for addressing these relatively simple questions.

As discussed in Section II, the flux tube model is a model of nonrelativistic quarks moving in the potential (\ref{ipv}), which incorporates the effect of a phonon excitation on the flux tube connecting the quarks.  Because angular momentum can be exchanged between the flux tube and the quarks, the ground state of the Schr\"odinger equation is not an eigenstate of the quark quantum numbers $^{2S+1}L_J$.  Rather, for the $0^{+-}$ hybrid, it is a linear combination of $^3S_1$, $^3P_1$ and $^3D_1$ states.  In this model, $^3S_1$ state has the lowest energy, with the $^3P_1$ state excited by approximately $0.24\,{\rm GeV}$.  The analysis of Isgur and Paton~\cite{flux} finds a hybrid mass roughly in the middle, indicating significant mixing between the two states.  The $^3D_1$ configuration is excited by approximately another $0.2\,{\rm GeV}$ over the $^3P_1$ state; therefore its contribution to the ground state $0^{+-}$ hybrid is likely to be small, and we shall neglect it here.  There is no $^1S_0$ component in the one phonon sector, because in this model such states have the wrong charge under $CP$~\cite{flux}.  The decomposition (\ref{fock}) then takes the form
\begin{equation} \label{fock0}
  | 0^{+-} \rangle = A | c \bar c(^3S_1)_{1,8} + g_1 \rangle +
  B | c \bar c(^3P_1)_{1,8} + g_2 \rangle\,.
\end{equation}
In the case of ordinary quarkonium, it is possible to make a model independent identification of the leading (color singlet) Fock component, with the admixture of other components controlled by an expansion in powers of $v$.  For hybrids, the identification of a ``leading'' component is model dependent.  Here we will choose to truncate the Fock expansion at these two quark configurations, and, for reasons discussed in Section IIIA, we will take the quarks to be in a relative color octet.  However, the constants $A$ and $B$ will remain free parameters, subject to the constraint $|A|^2+|B|^2=1$.

To calculate the desired matrix elements in this model, we need wavefunctions which are regular as $r\to 0$.  Hence we will solve the Schr\"odinger equation separately for the states of definite quark orbital angular momentum $L$, and use these solutions to estimate matrix elements such as $\langle O_8(^3S_1)\rangle$ and $\langle O_8(^3P_1)\rangle$.  We neglect the direct contributions to the matrix elements from mixing between the eigenstates of $L$, which amounts to working only to first order in the phonon term in $V(r)$.  Within this model we can account explicitly for the coupling of the quarks to the short distance process, but not for the coupling of the flux tube to the soft gluon background.  We will parameterize by $\xi$, a nonperturbative parameter presumably of order one, the amplitude for the color octet current to excite the one phonon excitation on the flux tube.  We remind the reader that, within the present context, we have little insight into the actual magnitude of $\xi$, so much larger or smaller values than $\xi\sim1$ are also possible.

The flux tube, which itself carries color charge, has the effect of decorrelating the color states of the quarks at either end.  We will take a statistical approach, noting that an uncorrelated quark-antiquark pair has a probability ${8\over9}$ to be found in a relative color octet.  In addition, we have to project the total (spin and orbital) $J=1$ carried by the quarks and the unit angular momentum of the flux tube onto a hybrid of total spin zero.  This projection leads to another factor of ${1\over9}$ in the matrix elements.  Including these three factors, which combine to ${8\over81}\xi^2$, the matrix elements may now be written in terms of the nonrelativistic radial wave functions $R_S(r)$ and $R_P(r)$.  The result is~\cite{nrqcd}
\begin{eqnarray}
  \langle O_8(^3S_1)\rangle &=&{8\over81}\,\xi^2\,|A|^2\times{3\over2\pi}\,
  |R_S(0)|^2,\nonumber\\
  \langle O_8(^3P_1)\rangle &=&{8\over81}\,\xi^2\,|B|^2\times{9\over2\pi}\,
  |R'_P(0)|^2,\nonumber\\
  \langle P_8(^3S_1)\rangle &=&-{8\over81}\,\xi^2\,|A|^2\times{3\over2\pi}\,
  {\rm Re}\,\left[R^*_S(0)\,\mbox{\boldmath $\nabla$}^2 R_S(0)\right],
  \nonumber\\
  \langle Q_8(^3S_1)\rangle &=&-{8\over81}\,\xi^2\,|A|^2\times{1\over2\pi}\,
  {\rm Re}\,\left[R^*_S(0)\,\mbox{\boldmath $\nabla$}^2 R_S(0)\right].
\end{eqnarray}
There is a linear divergence in $\mbox{\boldmath $\nabla$}^2 R_S(r)$ as $r\to0$.  However, we may use the Schr\"odinger equation to rewrite $\langle P_8(^3S_1)\rangle$ as
\begin{eqnarray}
  \langle P_8(^3S_1)\rangle &=&{8\over81}\xi^2\,|A|^2\times{3\over2\pi}\,
  2m\left[V(0)-E\right]\,|R_S(0)|^2\nonumber\\
  &=&2m\left[V(0)-E\right]\,\langle O_8(^3S_1)\rangle\,.
\end{eqnarray}
Here it is clear that this divergence is proportional to $(\alpha_s/r)\langle O_8(^3S_1)\rangle$, so it is compensated by a linear divergence in the first radiative correction to the coefficient of $\langle O_8(^3S_1)\rangle$.  Since we have not included these radiative corrections in our operator product expansion, we will subtract this divergence explicitly~\cite{nrqcd}.  There is also a logarithmic divergence which we have subtracted implicitly, and which changes the matrix element by an amount of order one.  Since our primary purpose in computing $\langle P_8(^3S_1)\rangle$ is to show that it is small, this ambiguity is not a problem.  Solving the Schr\"odinger equation numerically, with potential parameters $b= 0.18\,{\rm GeV}^2$, $c=-0.7\,{\rm GeV}^2$ and $f=1$, and a constituent charm quark mass of $1.77\,{\rm GeV}$, we find
\begin{eqnarray}
  \langle O_8(^3S_1)\rangle &=&{4\over27\pi}\xi^2\,|A|^2\,
  \left[0.58\,{\rm GeV}^3\right]\,,\nonumber\\
  \langle O_8(^3P_1)\rangle &=&{4\over9\pi}\xi^2\,|B|^2\,
  \left[0.03\,{\rm GeV}^5\right]\,,\nonumber\\
  \langle P_8(^3S_1)\rangle &=&{4\over27\pi}\xi^2\,|A|^2\,
  \left[0.14\,{\rm GeV}^5\right]\,,\nonumber\\
  \langle Q_8(^3S_1)\rangle &=&{4\over81\pi}\xi^2\,|A|^2\,
  \left[0.14\,{\rm GeV}^5\right]\,.
\end{eqnarray}
Since in this model $\langle O_8(^1S_0)\rangle=\langle P_8(^1S_0)\rangle=0$, 
the expansion (\ref{answer}) becomes
\begin{equation}
  \Gamma \simeq\Gamma_0\,
  {16(1-\mu_\psi)^2(1 + 2 \mu_\psi )\over 27\pi m_b^2 m_\psi}\,
   \xi^2\left[0.58|A|^2+0.10|B|^2+0.08|A|^2-0.03|A|^2+\dots\right],
\end{equation}
where we have taken $m_c=1.35\,{\rm GeV}$.  The second, third and fourth terms are formally suppressed by $v^2$ compared to the first, and we see that within this model they are indeed relatively small.  Taking $m_\psi=4\,{\rm GeV}$ and $m_b=4.8\,{\rm GeV}$ we find
\begin{equation}
  \Gamma(B\to\psi_g+X)\simeq(2\times 10^{-4})\,\xi^2\,\Gamma_0\,.
\end{equation}
The inclusive rate $\Gamma(b\to c\bar cs)$ is suppressed relative to $\Gamma_0$ by a phase space factor of approximately 0.25.  Taking $A=1$, to be conservative, and the amplitude for flux tube creation $\xi\alt1$, we obtain
\begin{equation}
  \Gamma(B\to\psi_g+X)\alt(1\times 10^{-3})\,\Gamma(b\to c\bar cs)\,.
\end{equation}
According to this estimate, then, the production of this hybrid charmonium is more than a factor of ten lower than that of typical ordinary charmonium states.  Of course, there is considerable model dependence in this calculation, and we have considered the production only of a single hybrid.  Nonetheless, given the tiny overall magnitude of the result, we conclude that it is quite unlikely that this mechanism could play an important role in ``hiding'' the charm produced in $b\to c\bar cs$ transitions.  The addition of additional hybrid channels will not change this qualitative conclusion.

\section{Tensor charmonium production in B-decays}

The decay into $D\overline D$ pairs is prohibited by parity not only for some exotic hybrids, but also for the ordinary $D$ wave charmonium states
$\psi(2^{--})$ and $\psi(2^{-+})$ (respectively, the $\bar cc$ in a $^3D_2$ and a $^1D_2$ configuration).  Perhaps the production of these tensor mesons, and their subsequent decay to light hadrons, could also ``hide'' the charm produced in underlying $b\to c\bar cs$ transitions.  The decay rates for $B \to \psi(2^{- \pm}) + X$ have been computed using the same NRQCD techniques developed for conventional quarkonium in Ref.~\cite{ko}, with the results
\begin{eqnarray} \label{tensor}
  \Gamma (B \to \bar cc(^3D_J) + X) &=&
  \Gamma_0\, {2(1 + 2 \mu_\psi)(1-\mu_\psi)^2\over m_b^2 m_c}
  \left [
  \langle O_8^{^3D_J}(^3S_1) \rangle + \frac{6}{m_c^2}
  \langle O_8^{^3D_J}(^3P_0) \rangle \right ], \nonumber \\
  \Gamma (B \to \bar cc(^1D_2) + X) &=&
  \Gamma_0\, {6(1-\mu_\psi)^2\over m_b^2 m_c}\,
  \langle O_8^{^1D_2}(^1S_0) \rangle\,,
\end{eqnarray}
where the approximation $m_\psi=2m_c$ was made.  Note that coefficients of the operators $\langle O_8(^3S_1)\rangle$ and $\langle O_8(^1S_0)\rangle$ are the same as in Eq.~(\ref{answer}).  Using velocity scaling rules and data from $Z^0$ decays, the corresponding branching fractions were
estimated to be ${\cal B} (B \to \psi(2^{--}) + X) \simeq 1.1 \div 5.2 ~\%$, and ${\cal B} (B \to \psi(2^{-+}) + X) \simeq 0.7 ~\%$. 

However, for this mechanism
to be responsible for boosting $\Gamma(b \to {\rm no\ charm})$,
the branching ratios of $\psi(2^{--})$ and $\psi(2^{-+})$
to light hadrons must be sufficiently large.
Unfortunately, this does not appear to be the case here.
For the $\psi(2^{--})$, one estimates
${\cal B} (\psi(2^{--}) \to \chi_{c1} + \gamma) \simeq 0.64$,
${\cal B} (\psi(2^{--}) \to \chi_{c2} + \gamma) \simeq 0.15$, and
${\cal B} (\psi(2^{--}) \to J/\psi \pi \pi) \simeq 0.12$~\cite{qiao}. 
Assuming that
the rest of the decays are into light hadrons, we obtain
${\cal B} (\psi(2^{--}) \to {\rm light\ hadrons}) \simeq 0.09$ and the combined
branching fraction
${\cal B} (B \to \psi(2^{--}) + X) \times {\cal B} (\psi(2^{--}) \to {\rm light\ hadrons})
\simeq (1 \div 5) \times 10^{-3}$. The addition of the
$\psi(2^{-+})$ does not change the situation: taking
${\cal B} (\psi(2^{-+}) \to h_c (^1P_1) + \gamma) \simeq 0.8$ as the
dominant decay mode and assigning the rest to
${\cal B} (\psi(2^{-+}) \to {\rm light\ hadrons})$ results in the combined branching fraction
${\cal B} (B \to \psi(2^{-+}) + X) \times {\cal B} (\psi(2^{-+}) \to {\rm light\ hadrons})
\simeq (1 \div 2) \times 10^{-3}$.  It is clear that unless the estimates of
NRQCD matrix elements are extremely inaccurate, this mechanism cannot play an important role in enhancing the charmless branching fraction in $B$ decays.

\section{Conclusions}

In the spirit of nonrelativistic quantum mechanics, we have developed a QCD-based expansion for the production of hybrid charmonium in $B$ meson decays.  Although no controlled Fock state expansion exists for hybrids, even in the limit of large $m_c$, our treatment provides a useful phenomenological starting point for model calculations of hybrid matrix elements.  In particular, it allows us to organize the various contributions to the production process in powers of the charm quark velocity $v$, since in most models the quarks are, in fact, nonrelativistic for large $m_c$.  Although we have treated explicitly only production in $B$ decays, the same matrix elements which appear here will also govern other production processes, as well as decays of hybrid states.

Based on the nonrelativistic expansion, we have used the flux tube model of Isgur and Paton to estimate the rate for the production of the exotic $J^{PC}=0^{+-}$ charmonium hybrid in $B$ decays, and found that it is considerably smaller than the production of ordinary charmonium.  Hence it is quite unlikely that this mechanism can play a significant role in resolving the difficulties with the semileptonic branching ratio and charm multiplicity observed in some experiments.  We stress that we have employed a very simple model, with the goal of obtaining an order-of-magnitude estimate only.  As models of hybrids become more refined, and as lattice studies develop further, more reliable calculations based on this expansion will become possible.

Finally, we have restricted ourselves to an estimate of the production of the $0^{+-}$ exotic hybrid, because its decays to $D^{(*)}D^{(*)}$ are prohibited by spacetime symmetries and it is likely to be among the lightest charmonium hybrid states.  Nevertheless, its mass is not yet known, and it is certainly possible that other hybrids, perhaps even ones with conventional quantum numbers, may be more important in $B$ decays.  This is a question which could be addressed with a more comprehensive and up-to-date survey of available models, but we would prefer to wait until the spectrum of hybrid charmonium, and the decays of the various states, have been revealed by experiment.

\acknowledgments

We gratefully acknowledge conversations with Steve Godfrey, Nathan Isgur, Philip Page, Eric Swanson and Adam Szczepaniak.  This work was supported in part by the United States National Science  Foundation under Grant
No.~PHY-9404057. A.F.~was also supported by the United States National Science
Foundation under National Young Investigator Award No.~PHY-9457916, by the United
States Department of Energy under Outstanding Junior Investigator Award No.~DE-FG02-94ER40869, by the Alfred P.~Sloan Foundation, and by the Research Corporation as a Cottrell Scholar.

\end{document}